\documentclass[aps,prl,reprint,superscriptaddress,10pt,%
floatfix,notitlepage]{revtex4-1}
\usepackage{graphicx,color}
\usepackage{amsmath,amssymb,nicefrac}
\usepackage{natbib}
\usepackage[utf8]{inputenc}
\usepackage{microtype,hyperref}

\let \vec \boldsymbol

\renewcommand{\exp}[1]{\mathrm{e}^{#1}}
\let \phi \varphi 
\let \theta \vartheta
\newcommand{\mat}[1]{\mathsf #1}

\begin{document}

\author{Horst-Holger Boltz}
\affiliation{James  Franck Institute, University  of  Chicago, Chicago,  IL  60637,USA}

\author{Jorge Kurchan}
\affiliation{Laboratoire de Physique Statistique, D\'epartement de physique de l'ENS, \'Ecole Normale Sup\'erieure,
PSL Research University; Universit\'e Paris Diderot, Sorbonne Paris-Cit\'e; Sorbonne Universit\'es,
UPMC Univ. Paris 06, CNRS; 24 rue Lhomond, 75005 Paris, France}
\author{Andrea J. Liu}
\affiliation{Department of Physics and Astronomy, University of Pennsylvania, Philadelphia, PA 19104, USA}

\title{Fluctuation Distributions of Energy Minima in Complex Landscapes}
\begin{abstract}We discuss the properties of the distributions of energies of minima obtained by gradient descent in complex energy landscapes. We find strikingly similar phenomenology across several prototypical models. We particularly focus on the distribution of energies of minima in the analytically well-understood $p$-spin-interaction spin glass model. We numerically find non-Gaussian distributions that resemble the Tracy-Widom distributions often found in problems of random correlated variables, and non-trivial finite-size scaling. Based on this, we propose a picture of gradient descent dynamics that highlights the importance of a first-passage process in the eigenvalues of the Hessian. This picture provides a concrete link to problems in which the Tracy-Widom distribution is established. Aspects of this first-passage view of gradient-descent dynamics are generic for non-convex complex landscapes, rationalizing the commonality that we find across models. 
\end{abstract}

\maketitle

\section{Introduction}
 
The notion of an underlying complex energy landscape in glassy, disordered systems is useful
~\cite{goldstein,stillinger,wales,onuchic,heuer1997,*heuer2008,krzakala2007,berthier2011,charbonneau2014} to the extent that the landscape can be reduced to relatively few properties that are relevant to observed phenomena. The complexity, which counts stationary points in the landscape (minima, saddles, maxima) is an example of such a property. An energy landscape is complex if the number of stationary points depends exponentially on the system size. 

An intuitive approach to probing complexity is to do a naive search for minima using gradient descent.~\cite{numrec} One follows an initial configuration along the (negative) gradient flow of the energy until a stationary point (vanishing gradient) is found. Because a numerical descent almost certainly ends in a minimum, gradient descent does not only constitute the simplest form of physical dynamics in a complex landscape, a quench to zero temperature, but also the most intuitive and simplest form of optimization. If one starts with flatly sampled random initial positions (corresponding to infinite-temperature $T=\infty$ configurations), gradient descent has the added advantage of sampling local minima with a probability that can be calculated because it is proportional to the volumes of their basins of attraction~\cite{xu2011,frenkel2017}.   Finally, in addition to being a local optimization strategy, gradient descent is also the archetypal greedy algorithm, particularly if one considers a discretized version as one does with any numerical implementation: in every time step the displacement with the largest expected loss in energy is chosen. Within the field of glassy systems, gradient descent is used to obtain ``inherent structures''~\cite{stillinger1984,*stillinger1985,sciortino1999,sastry2001, debenedetti2001}, \emph{i.e.} the minima at the bottom of the local basin of attraction around which the system thermally fluctuates, while in machine learning, gradient descent is the original go-to learning strategy~\cite{lecun2015}. Gradient descent is also used to obtain jammed packings of repulsive soft spheres, which are the least stable packings that are mechanically rigid~\cite{ohern2003,goodrich2014}.

Here we look at the shape of the distribution of minima obtained by gradient descent for several different models, with particular focus on the spherical $p$-spin-interaction spin glass. Such distributions, for example for jamming, have been assumed to be Gaussian~\cite{ohern2003}. Our central finding is that for all of these models, the distributions are non-Gaussian with non-trivial tail exponents on one side that are consistent with the Tracy-Widom distribution, a distribution mostly known for describing the edge fluctuations of the eigenvalues of Gaussian random matrices. We rationalize this finding with a novel perspective that might be the starting point for an eventual analytical approach.

In Sec.~II we introduce the models studied. We then present our numerical results in Sec.~III and use established results for the $p$-spin model in Sec.~IV to formulate a toy process that allows us to understand these numerical results. We close in Sec.~V with some final remarks on the applicability of these ideas to other contexts.

\section{Models \& Complexity}

We study various models with complex landscapes. A unifying perspective is provided by all of them being random constraint-satisfaction problems, i.e. assemblies of equations or inequalities. Generically, the question of interest is whether a specific realization allows for an assignment of the variables that satisfies all constraints or whether there are frustrations (which are easily introduced in randomized problems) that prevent satisfaction of all of the constraints. Generically there is a (SAT/UNSAT) transition between a phase where a satisfying assignment is possible (SAT) and a phase where this is not possible (UNSAT) upon tweaking the hardness of the satisfaction problem, e.g. by changing a control parameter such as the ratio of (in-)equalities and variables. Versions with discrete (particularly Boolean) variables are of fundamental importance to computer science\cite{cook1971}, whereas SAT/UNSAT transitions in continuous constraint-satisfaction problems are conjectured to form an important universality class~\cite{franz2017} in statistical physics. The focus of our attention is the spherical $p$-spin model which we therefore introduce first, before the $k$-SAT, perceptron and jamming models. 

\paragraph{The $p$-spin model.} Specifically we consider the spherical $p$-spin model\cite{crisanti1992,kurchan1996}: \emph{i.e.}, we have $N$ spins $S_i$ whose combined length is constrained to $\sum S_i^2=N$ (leaving effectively $N-1$ degrees of freedom) with an energy functional
\begin{align}
H &= \sum_{i_1<i_2<\ldots<i_p} J_{i_1,i_2,\ldots,i_p} S_{i_1} S_{i_2} \ldots S_{i_p}
\end{align}
containing random Gaussian couplings $J$ with mean zero and variance $\langle J^2\rangle_c = N/\#J$.
Here, $\#J \sim N^p$ is the number of terms appearing in the energy functional (while adhering to the constraint of ascending indices). We use this convention to account for finite-size effects from lower-order terms, but ultimately only the scaling with $N$ is important. Note that particularly in the older physics literature a different convention is used that introduces an additional factor of two here. This energy is an extensive quantity scaling with system size and we therefore also introduce the corresponding intensive quantity $\varepsilon = E/N$.
 As the qualitative nature of the energy landscape defined by this functional is independent of $p$ for $p>2$ ($p=2$ corresponds to a convex eigenvalue problem and therefore only has a single, trivially global minimum), we choose to limit ourselves to the numerically most accessible case of $p=3$. Still, the cost of a simple evaluation of the energy inevitably scales as $N^p$.
 
The energy scale $\varepsilon_{th}=-2\sqrt{(p-1)/p}=-\sqrt{8/3}$ is called the threshold energy as it constitutes the upper energy boundary below which an exponentially large number of stationary points exist. This is quantified by looking at the (cumulative) complexities. If we define $\mathcal{N}_k(\varepsilon)$ to be the number of stationary points of index $k$ with an (intensive) energy not larger than $\varepsilon$, the corresponding complexity $\Sigma(\varepsilon)$ is given by
\begin{align}
\Sigma(\varepsilon) &= \frac{1}{N} \log \mathcal{N}_k(\varepsilon) \text{.} \label{eq:complex}
\end{align}
The complexity was studied earlier within the TAP approach~\cite{crisanti} and has been the subject of rigorous mathematical analysis in the limit of large $N$~\cite{auffinger}. Remarkably, a qualitatively similar structure has been found for rather small system sizes by numerical enumeration of the critical points~\cite{mehta}. 

In this paper, we focus on the shape of the distribution of energies of minima, as obtained by gradient descent for finite systems. This corresponds to the shape of the normalized distribution corresponding to $\mathcal{N}_{k \equiv 0}(\varepsilon)$. The distribution of final energies found as a result of gradient descent for the $p$-spin model is shown in fig.~\ref{fig:combo_pdf} (a).

For a suitable choice of couplings, the $p$-spin model provides a natural energy landscape for the optimization problem corresponding to a $k$-SAT decision problem~\cite{mezard2002}. The model also provides insight into structural glasses~\cite{cugliandolo1993}. It is also a valuable model in its own right.  The overall gestalt of the energy landscape, as captured by the complexities, is the relevant property that drives interest in the $p$-spin model as a prototypical complex energy landscape. Physical systems usually have a well-defined notion of a ground-state energy which sets a lower bound to an extensive number of minima. Additionally, the existence of an upper bound reflects that ``over-frustration'' of a complex system--it is exponentially hard to construct a state with an energy less favorable than some native scale. 

\paragraph{The $k$-SAT model.} The prototypical satisfiability problem is that of Boolean (or propositional) satisfaction, see for example ref.~\citenum{gogioso} for an introduction. Given a number $N$ of literals (Boolean variables $s_i$ with $s_i \in \{ \text{\tt TRUE}, \text{\tt FALSE}\}$) and a number $M$ of clauses (combinations of the literals and the fundamental logical operators {\tt OR} ($\lor$), {\tt AND} ($\land$) and {\tt NOT} ($\neg$)) which can always be brought into conjunctive normal form, which means that we consider conjunctions (AND-connected sub-clauses) of disjunctions (OR-connected (possibly negated) literals), the goal is to find a choice of the literals that satisfies the clauses (evaluates to a true statement). Particularly, we focus on the $k$-SAT version of this problem consisting only of random clauses that are disjunctions of exactly $k$ (possibly negated) literals. Interestingly, it turns out that there is a sharp change in the difficulty of the problem with $k$. For $k\leq 2$, the solution (or the existence of a solution) can be found easily in a time that depends polynomially on the problem size (see, for example, \cite{krom}), whereas the problem is NP-hard\cite{karp} for larger values of $k$ (for efficiency reasons we limit ourselves to $k=3$), meaning that the question of the existence of such an algorithm with polynomial runtime is an important outstanding problem\cite{fortnow}. Here, we are not interested in designing a particularly good algorithm. Instead, in analogy to gradient descent, we use a local greedy optimization strategy: pick an unsatisfied clause and an undetermined literal and set the literal to the value satisfying the clause. If at some point there are no undetermined literals left to satisfy an unsatisfied clause, the system is considered unsatisfied; if every clause is satisfied, the system is considered satisfied. The variable controlling the fraction of unsatisfied systems is $M/N$, the ratio of the number of clauses to the number of literals (the solution is obviously trivial if every literal appears in at most one clause). Because the literals are Boolean the distribution of results is not continuous, but the relevant combination $M/N$  becomes continuous in the thermodynamical limit and we will, thus, treat the data as if they were binned continuous data. We perform runs of the greedy algorithm for $k=3$ (ensuring that every literal is used in any clause at most once) with $N=128$ literals for $M=1\ldots 10^3$ and measure the fraction of unsatisfied systems. This data is shown in fig.~\ref{fig:combo_pdf} (a).

Reduction from $k$-SAT is usually used to prove that other Boolean satisfaction or decision problems are NP-hard; for example, there is a direct connection between the 3-Sat and the 3-coloring of a graph (by means of factor graphs). The optimization problem associated with the $k$-SAT problem is the $p$-spin model discussed above.

\paragraph{The perceptron model.} Generalizing from Boolean to continuous variables, there are two types of constraints: equality constraints ($f(\vec S)=0$) and inequality constraints ($f(\vec S)\geq 0$). Every independent equality constraint reduces the dimension of the solution space by one, meaning that the set solutions to problems with equality constraints is always one of zero measure geometrically, i.e. the solutions are isolated in the space of possible solutions. This is not only peculiar, but it is also inappropriate for many, if not most actual systems one might want to model: a very simplified descriptions of neurons, for example, is that they give an output if the input is exceeding some threshold (see jamming below for another example). This model\cite{mcculloch} of neurons is the origin of the perceptron model~\cite{gardner1988}. Following the notation of Ref.~\citenum{franz2015}, we consider continuous variables $S_i$ ($i=1...N$) subject to linear inequality constraints $h_\mu$ ($\mu=1...M$) such that $h_\mu = \vec \xi_\mu \cdot \vec S - \sigma_\mu \stackrel{!}{>} 0$. We limit the variable space to a sphere, $\vec S^2 = N$, use normally distributed $\vec \xi$ and set $\sigma_\mu \equiv \sigma$. For a given ratio $\alpha=M/N$ there is a critical value $\sigma_c(\alpha)<0 $ that marks the satisfiability transition (lower $\sigma$ corresponding to the phase in which all constraints are satisfied). Tuning $\sigma$ allows for further control of the topology (convex/non-convex) of the energy landscape that is constructed by considering $E(\vec S) = \sum_\mu h_\mu^2 \Theta(-h_\mu)$, transforming the decision into an optimization problem. At low values of $\sigma$ the system is convex and at sufficiently high values (but for $\sigma>\sigma_c$) the system is non-convex. We perform gradient descents on that energy landscape to obtain the distributions of the final energies in both the non-convex (Fig.~\ref{fig:combo_pdf}(c)) and convex regimes (Fig.~\ref{fig:combo_pdf}(d)).

\paragraph{Jamming.} We consider the packing\cite{ohern2003,liu2010} of spheres with harmonic repulsions in low ($d=2,3$) dimensions. The specifics of the interaction potentials (e.g. using anharmonic spheres) affect the nature of the transition as captured by critical exponents~\cite{ohern2003}, but do not change the underlying SAT-UNSAT motif. To see this, one simply scales the energy by an effective spring constant that depends on the interaction potential and on pressure or average coordination number; once this is done, the results for different potentials collapse. We therefore study only the case of harmonic repulsions for simplicity. Starting from randomly placed spheres, the energy is lowered by reducing the overlap of the spheres; the constraints to be satisfied are, thus, of the form $\lvert \vec{r}_{ij} \rvert \geq \sigma_{ij}$ where $\vec{r}_{ij}$ denote the pairwise displacement vectors between particles and $\sigma_{ij}$ are the added particle radii. As a result of the inequality structure of these constraints, the perceptron model provides the appropriate mean-field framework \cite{franz2015,franz2016}. The relevant control parameter that sets whether or not an unjammed configuration (which we choose to have $E=0$) is found is the packing fraction, i.e. the ratio of combined volume of the spheres to volume available in the simulation box. For finite dimensions and particle sizes, there is not a sharp satisfiability threshold, but as one increases the packing fraction the fraction of systems for which the descent ends in a jammed configuration increases. The derivative of the satisfaction curve can be interpreted as the distribution of jamming thresholds $\phi_c$ and is shown (as inferred via numerical derivation from the data of ref.~\citenum{vagberg}) in Fig.~\ref{fig:combo_pdf} (f). Previously, these distributions were used to infer finite-size scaling properties\cite{ohern2003,vagberg} such as the scaling of the width of the distribution and of the jamming threshold value in the thermodynamic (large system size) limit. In contrast, we focus on the shape of these satisfiability distributions.

Additionally, we consider the distribution of energies of packings (using the same model as e.g. ref.~\citenum{ohern2003} with $\alpha=2$) prepared\cite{ridout} at a fixed pressure above the jamming transition, where not all of the constraints are satisfied. These curves are shown in Fig.~\ref{fig:combo_pdf}(e).

\section{Numerical Results}

\begin{figure*}
\includegraphics[width=1.0\linewidth]{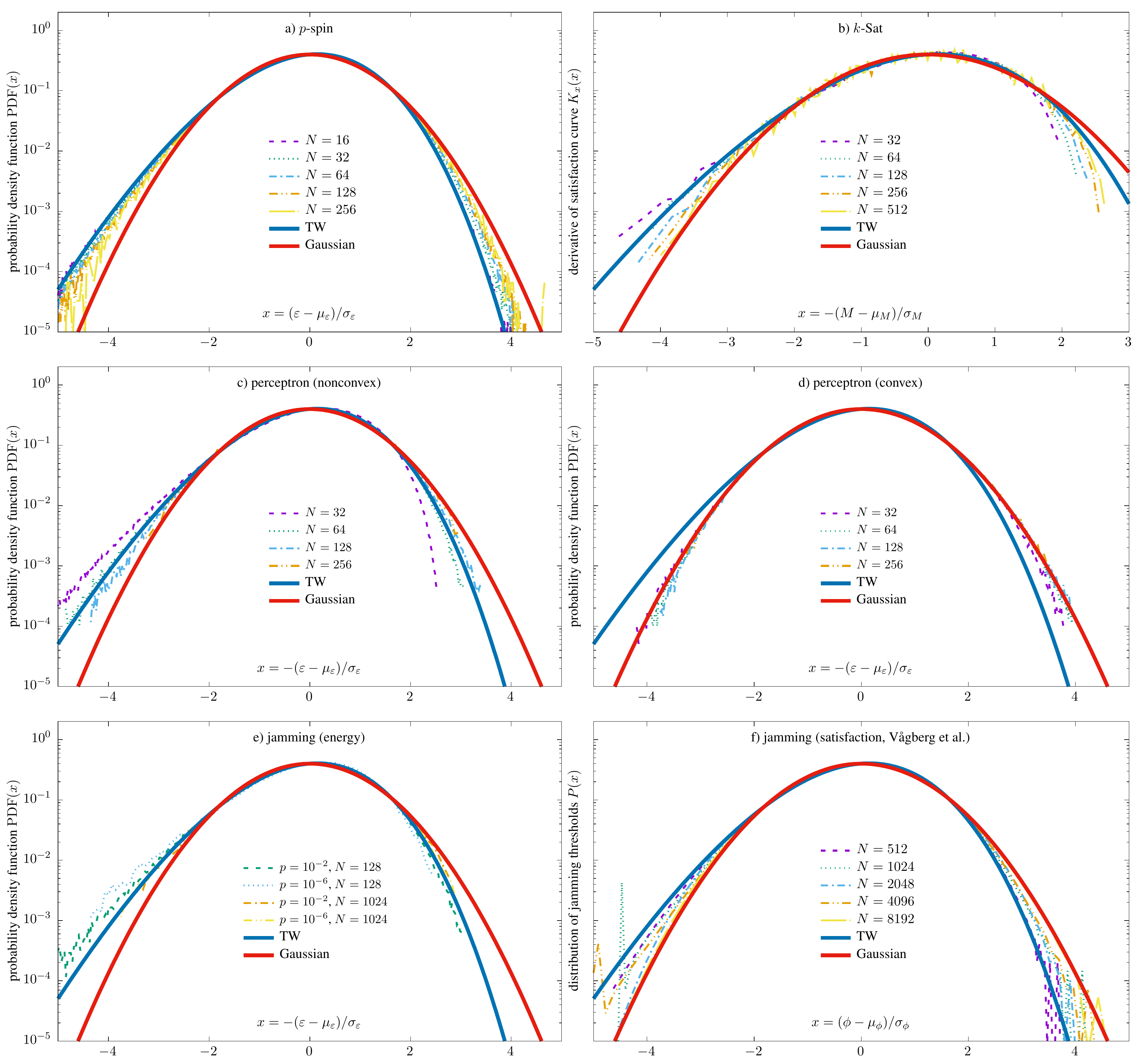}
\caption{Fluctuation distributions for the models discussed in the main text as found from numerical descents. Each panel also contains the curve corresponding to data from the Tracy-Widom distribution $W$  (analogously normalized) and the Gaussian normal distribution. {\bf a)} Derivative of the fraction of satisfied systems in the $k$-Sat model as a function of the number of clauses $M$. Here, normalization was performed as if the satisfaction curve were a cumulative distribution. {\bf b), d)} Energy distributions for the perceptron ($\alpha=5$; curves in b) labeled with $\sigma=\ldots$), isobaric jamming of soft spheres (curves in b) labeled with $p=\ldots$) and the $p$-spin (d)). {\bf c)} Distribution of jamming thresholds $\phi_c$ found in the jamming of soft spheres according to ref.~\protect\citenum{vagberg}. The binning size is determined following ref.~\protect\cite{shimazaki}.}
\label{fig:combo_pdf}
\end{figure*}

We present the results of gradient descent simulations for all the models studied in Fig.~\ref{fig:combo_pdf}. Because shifting and global rescaling of the energy landscape do not qualitatively affect gradient descent, we only present histograms of normalized variables (mean zero, unit variance). The bulk of these simulations was done employing the FIRE algorithm\cite{bitzek} instead of a naive, direct integration of the equation of motion. This algorithm converges significantly faster, allowing for better statistics and analysis of large deviations. The additional inertial degree of freedom within the FIRE scheme can in some individual cases change the basin of attraction such that the relaxation from a specific initial condition with it leads to a different final minimum than would application of a direct gradient descent  (this is also true for gradient descents with different time steps). However in smaller runs, we find no indication that this changes the statistics significantly, hinting that the effect of the additional inertial movement within FIRE mostly changes the time axis. This is in line with previous applications of FIRE in similar quenches to zero temperature in jamming. While FIRE has seen greater usage within the jamming model, we are using it for all the continuous problems.

As a visual aid and for comparison, we show the respective numerical probability functions alongside two distributions: (1) the Gaussian (normal) distribution, which is the least biased estimator for the distribution having fixed the mean and variance, and (2)  the (normalized) Tracy-Widom distribution, which is characterized by tails that decays more slowly than a Gaussian on one side ($x \ll 0$) and more rapidly than a Gaussian on the other side ($x \gg 0$):
\begin{align}
\log W (x) \sim \begin{cases} x^{3/2} & x \ll 0 \\ -x^{3} & x\gg 0 \end{cases}. \label{eq:tw}
\end{align}
There is a striking qualitative similarity across models and system sizes. The distribution functions are trivially similar to the Gaussian around zero, but the large deviations are asymmetric with a soft tail (in our presentation for $x<0$) that decays more slowly than the Gaussian and a hard tail (for $x>0$) that decay more rapidly than the Gaussian. The soft tail seems to be well-described by the Tracy-Widom distribution. The strong commonality across systems and the Tracy-Widom form of the soft tail constitute the main results of this paper. There  is some additional $N$-dependence not eliminated by normalizing, which partially is to be expected for small systems due to corrections to scaling ( see Ref. \cite{goodrich2016} for a discussion in the specific context of jamming; we note that corrections to scaling are expected for all of the systems studied). Nevertheless, the soft tail appears robust in the thermodynamic limit, as we will elaborate below. Interestingly, the notable exception to this is the perceptron in the convex regime that has a trivial basin of attraction. This indicates that the important similarity between the analogous systems is indeed the quench from a flat measure in a complex energy landscape and the exploration of at least partially concave (some eigenvalues of the energy landscape are negative) regions.

\begin{figure}
\includegraphics[width=\linewidth]{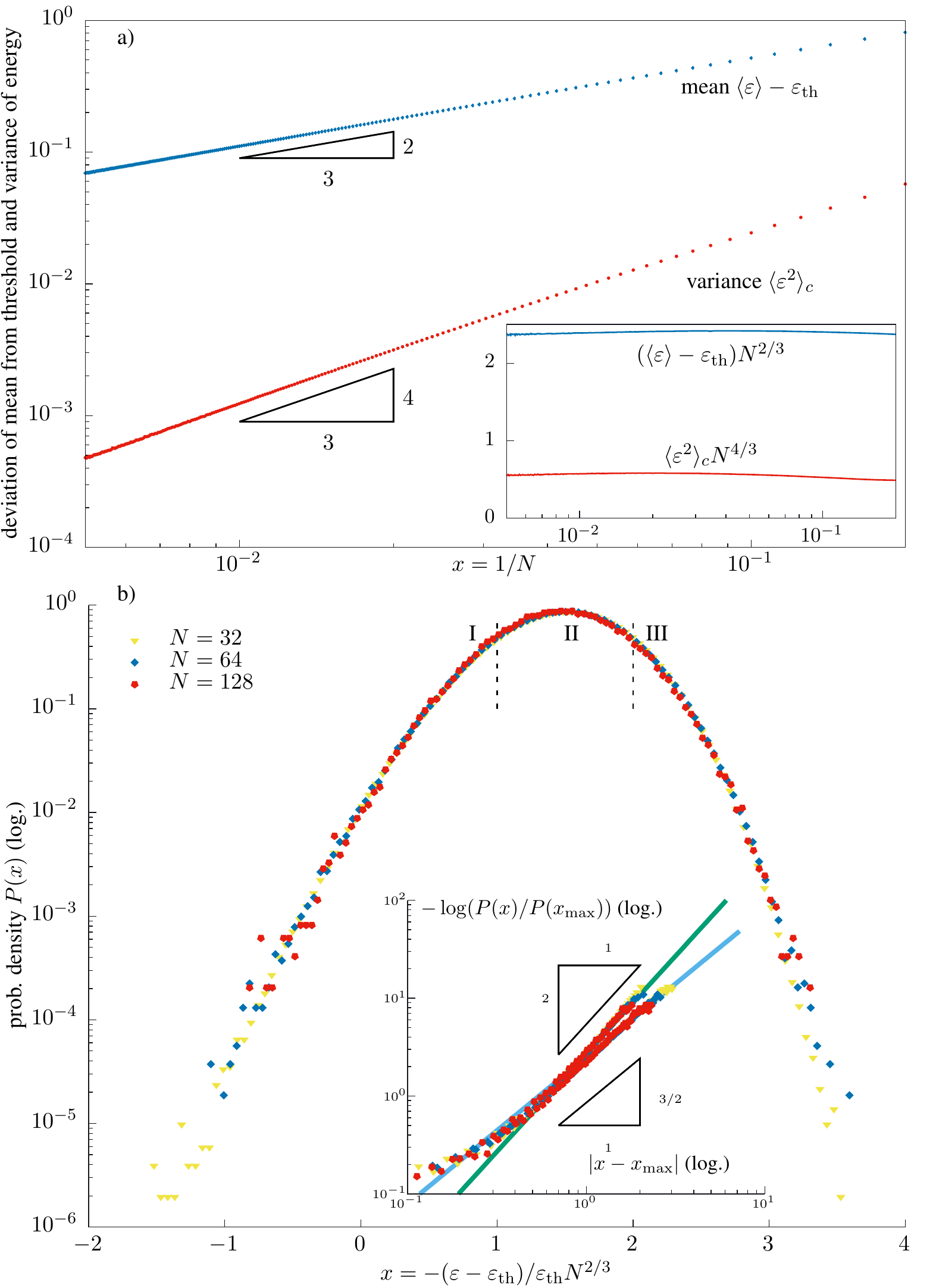}
\caption{a) Mean and variance of the final energies as found via steepest descent in the $p$-spin as a function of system size $N$ (double logarithmic plot). We find that the energy clearly descents towards the threshold energy from above with a very clear power-law dependence (see inset). b) Collapsed distributions for $N=32,64,128$. Here, we perform collapsing using a tentative non-normalized scaling form that is inspired by the Tracy-Widom distribution. The labels I,II and III indicate the three regions used for the spectral densities in Fig.~\protect\ref{fig:ev}. Inset: Large deviations, power-laws are found at the far tails of the energy distribution. }
\label{fig:fs_distro}
\end{figure}

We now look carefully at finite size effects to see whether the soft tail survives in the limit $N \rightarrow \infty$. We present results for the spherical $p$-spin model, for which we have the best statistics {($4\cdot 10^7$ samples for $N\leq 64$, $7\cdot 10^6$ for $N=128$ and still $80000$ for $N=256$)  and which is also expected to have very small corrections to scaling  due to its structure. This is highlighted by the extremely clear power-laws found for the first two moments of the final energies (see Fig.~\ref{fig:fs_distro}(A)). Using overlines to denote averages over  gradient descent samples (so as not be confused with unbiased averaging over disorder for which we use angular brackets), we find that the finite-size deviations to the energy can be characterized via
\begin{subequations}
\begin{align}
\overline{\varepsilon}-\varepsilon_\text{th} &\sim N^{-2/3}\\
\overline{\varepsilon^2}_c & \sim N^{-4/3} \text{.}
\end{align}
\label{eq:fs-law}
\end{subequations}
We denote cumulants with a subscript, such that $\overline{x^2}_c=\overline{x^2}-\overline{x}^2$.

Studying the large deviation tails, we find that the ``soft'' tail (corresponding to low energies) has the same asymptotics as in the Tracy-Widom law, but the ``hard'' tail decays more like a Gaussian and therefore decays considerably more slowly than the Tracy-Widom law,
\begin{align}
\log P_\text{empirical} (x) \sim \begin{cases} x^{\approx 3/2} & x \ll 0 \\ -x^{\approx 2} & x\gg 0 \end{cases}
\end{align}
with again the rescaled variable $x=(\varepsilon-\mu_\varepsilon(N))/\sigma_\varepsilon(N)$. This is shown for the $p$-spin model in Fig.~\ref{fig:fs_distro}(B). From Fig.~\ref{fig:combo_pdf} (a) it is hard to tell whether this holds for large $N$ as the shape of the distribution appears to cross over from something close to the Tracy-Widom distribution towards the Gaussian distribution. To address this question, we construct an estimate for the converged shape for very large $N$ using the following procedure. We sample the inverse function $Q_N(c)=C_N^{-1}(c)$ to the empirical cumulative distribution functions (CDFs) $C_N(x)$ for every $N$ (using the $N$-specific normalization) at a number of selected values $c=10^{-5},\ldots,1$. At any given value of $c$, this gives a set of pairs $(1/N,x_c=Q_N(c))$ which we use to extrapolate to $x_c(0)$. We find that the finite size effects in the shape are well described by $x(N)-x_c(0) \propto N^{-1/3}$ (corresponding to $1/N$ corrections in non-normalized variables). As the extrapolation is done at each value independently, this method does not constrain the moments of the final distribution and we therefore normalize it as a final step. The result of this procedure is shown in Fig.~\ref{fig:extrap}, which suggests that although the final distribution is extremely close to a Gaussian, the soft tail does prevail for large $N$.

\begin{figure*}
\includegraphics[width=\linewidth]{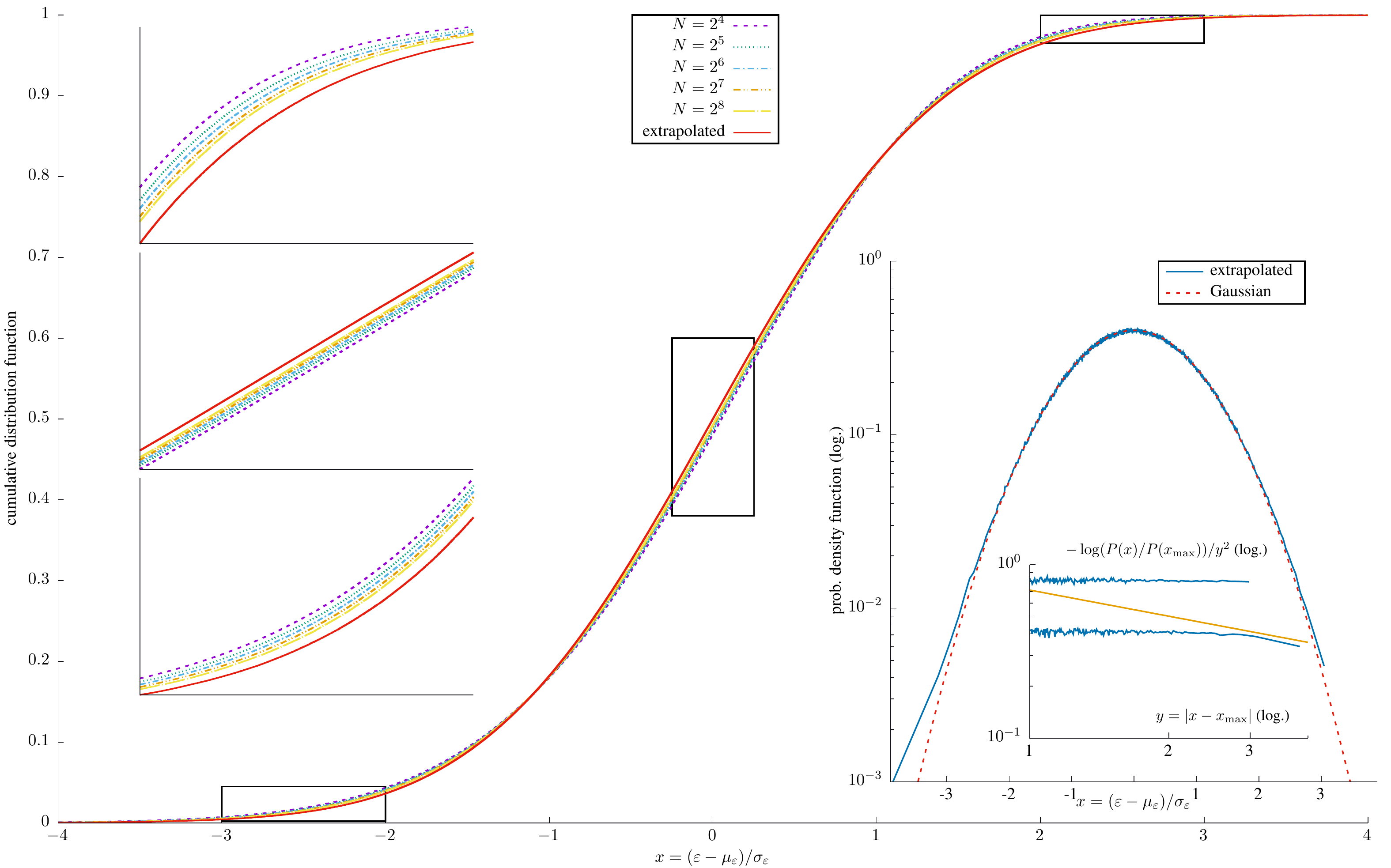}
\caption{Large: Cumulative distribution of final (normalized, intensive) energies in the $p$-spin model for $N=2^{4,..8}$ together with the extrapolated (refer to the main text for details) distribution for large $N$. Left: The convergence in the marked regions of the large plot is shown in insets. Right: The derivative of the extrapolated cumulative distribution, i.e. the extrapolated probability density function. The inset shows the tail behavior (relative to a Gaussian $y^2$-behavior) with a guide to the eye $\propto y^{-1/2}$ which corresponds to a $y^{3/2}$ soft tail for small energies. }
\label{fig:extrap}
\end{figure*}

\section{Rationalization of Results}

To gain insight into the finding that the distribution of minima energies is non-Gaussian, we shall construct a simplified meta-model which, we shall argue, seems to capture the essential elements of the dynamics. It is a first-passage problem for the lowest eigenvalue of a random matrix that with fluctuating elements that is being gradually shifted by an identity. 
We focus on the $p$-spin model.This is our starting model of choice because of its simple structure; we will show that we can learn something from the spectral dynamics here more clearly than in other models. In particular, we will relate the spectral dynamics to Dyson Browian motion. Such Dysonian dynamics are not general as we discuss in the final section, but reflect key elements that are true for the evolution of any dynamical matrix (noise, non-crossing of eigenvalues and entropical confinement). In this model, gradient descent is given by integration of\footnote{From here on out, we use a summation convention where repeated indices are to summed over. In the spirit of simplification, we also set $J_{ijk}=0$ unless $i<j<k$, wherever it is relevant.}
\begin{align}
\dot S_i &= -(J_{ijk}+J_{jik}+J_{jki}) S_j S_k - z S_i
\end{align}
where $z$ is a Lagrange parameter ensuring the the spherical constraint, $S_i S_i = N$, and is fixed to be
\begin{align}
z=-3 \varepsilon
\end{align}
by demanding $S_i \dot S_i = 0$ or, equivalently, $S_i S_i = \text{const}$. The descent terminates once $\dot S_i=0$ and this will be not only a stationary point, but a stable minimum. 

\begin{figure}
\includegraphics[width=\linewidth]{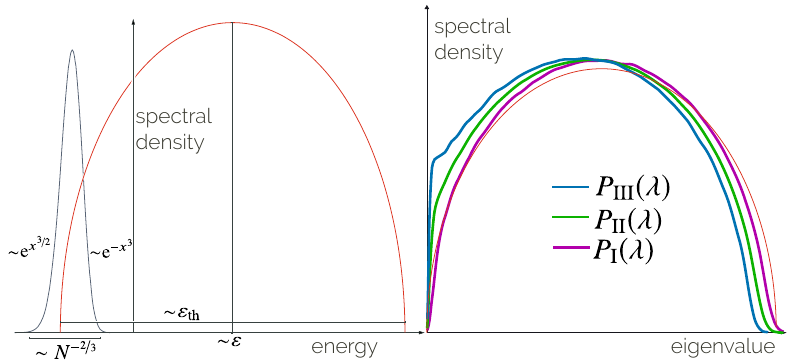}
\caption{Left panel: Schematic sketch of the Wigner semi-circle expected in a unbiased sampling from the Hessian. The location of the mean is set by the state's energy, the width corresponds to the threshold energy. The deviations from this average picture at finite system sizes are for the extremal eigenvalues given by a Tracy-Widom distribution with a characteristic finite-size scaling, $\sigma_{\lambda_\text{min}}\sim N^{-2/3}$. Right: eigenvalue distributions in different regimes (atypically low (I), typical (II), atypically high energy (II)), cp. Fig.~\ref{fig:fs_distro}. We show the marginal semi-circle law (thin red line)  as a visual guide.}
\label{fig:ev}
\end{figure}

To understand the implications of this, it is helpful to also consider the  dynamical matrix $\mat M$ associated with the descent, which is given by the Hessian of the energy function (technically the Lagrange function, but we refrain from reflecting the special nature of the spherical constraint in our wording in the following):
\begin{align}
M_{ij} &= (J_{ijk} + \ldots) S_k + z \delta_{ij} \label{eq:hessian}
\end{align}
where the omitted terms correspond to all index permutations of $i,j,k$. We can identify two contributions to the matrix $\mat M$: a Gaussian random part ($J$ and $S$ are practically independent) and a deterministic shift that only depends on the energy. From this observation, it is straightforward to infer that the spectral density of eigenvalues in the limit of large system sizes is given by a shifted Wigner semicircle law~\cite{wigner}
\begin{align}
\rho(\lambda) \,\mathrm{d}\!\lambda &= \frac{1}{2\pi\sigma^2} \sqrt{(2\sigma)^2 - (\lambda-\mu)^2}\,\mathrm{d}\!\lambda
\end{align}
with $\sigma=-3\varepsilon_{th}$ and $\mu = -6\varepsilon$, see left panel of Fig.~\ref{fig:ev}.

If we ignore the shift (the second term of the Hessian in Eq.~\ref{eq:hessian}) for now, then we know that finite system size causes the edges of the Wigner semicircle to develop fluctuations. It is rather intuitive that these fluctuations have to be asymmetric: finding a lowest eigenvalue that is smaller than the lower boundary of the support of the semi-circle should (for sufficiently large system sizes) be entropically less costly then finding a fluctuation where the lowest eigenvalue is located somewhere within the bulk of the semi-circle; this implies that an extensive number of eigenvalues must lie at atypically large values. This intuition has been made rigorous--the edge fluctuations are described by the acclaimed Tracy-Widom distribution introduced earlier in Eq.~\ref{eq:tw}~\cite{tracy1993,*tracy1994,deanmajumdar2006,nadal,*majumdar2014}. The Tracy-Widom distribution has gathered considerable interest in recent years as it has been found to appear in many systems of correlated variables that are beyond its original scope within random matrix theory. Interestingly, the charge-like repulsion between the eigenvalues that is at the core of the Tracy-Widom distribution is of purely topological origin and can be understood solely from imposing a non-crossing bias onto random walkers~\cite{grabiner}. As of now, there is no simple closed form representation of the Tracy-Widom distribution, but the distribution of the lowest eigenvalue has a characteristic system-size dependence\footnote{Note that our conventions are such that the spectrum is intensive. This results in different scalings then the also commonly used extensive spectrum where the width of the semicircle is of order $\sqrt{N}$.},
\begin{align}
P(\lambda_\text{min}=\lambda) \sim W(-(\lambda+2\sigma-\mu)\sigma^{-1} N^{2/3}) \label{eq:t-w-law}
\end{align}
with $W(x)$ being the Tracy-Widom distribution with tails described by Eq.~\ref{eq:tw}\footnote{The Tracy-Widom distribution is usually defined by the fluctuations of the largest eigenvalue, thus $W(x)=F_1(-x)$}.

Note that this eigenvalue spectrum has two tails with different $N$-scalings: in the soft tail the argument of the exponential scales like $N$ whereas it scales like $N^2$ in the hard tail. This corresponds to the fact that deviations of the lowest eigenvalue to smaller values are entropically suppressed by the definition of the matrix ensemble, but fluctuations to higher values require a displacement of extensively many eigenvalues.

The termination of the descent is subject to the gradient and, thus, cannot be understood by the dynamics of the eigenvalues alone, but we can identify a necessary contribution that will get us close to understanding the full dynamics. Once the lowest eigenvalue crosses zero to become positive, the descent is in its final valley and the energy will only change slightly. Neglecting this final part, gradient descent becomes a first-passage problem in the lowest eigenvalue. As time progresses and the system lowers its energy, the eigenvalues will move (with fluctuations) towards higher values while never crossing each other, until the lowest eigenvalue crosses zero so that all eigenvalues are positive. A direct empirical corroboration of the importance of eigenvalue fluctuations from the data, and a connection to problems usually connected to Tracy-Widom-laws, is that the power-laws seen in the finite size effects, see eq.~\eqref{eq:fs-law}, are consistent with the scaling seen in the Tracy-Widom law, cp. eq.~\eqref{eq:t-w-law}.

To start examining the descent from this spectral perspective, we calculate the spectrum of eigenvalues of the Hessian in three different ranges of energies of the minima in Fig.~\ref{fig:fs_distro}(B). In each of these ranges, the distribution is close to the semicircle expected at the threshold energy in unbiased sampling independent of the energy, with a shift that increases with energy. This is somewhat consistent with the finding that all the states found are close to threshold. However, they are always above the large $N$ threshold, $\varepsilon_\text{th}=-\sqrt{8/3}$, which means that the naively expected value for the lowest eigenvalue is negative and a large deviation is needed to constitute a mechanically stable state. Intuitively, the entropically least expensive way to do this is to aggregate all these eigenvalues closely above zero. This intuition has been made rigorous by an analysis by Dean and Majumdar (DM) \cite{deanmajumdar2008}. An important physical consequence of this aggregation around zero (forming an integrable singularity in the spectrum) would be an excess of very soft modes which is not only unphysical, but also completely contrary to the empirical findings in physical realizations of disordered systems in general or our data for the $p$-spin in particular. An immediate explanation for why such an aggregation of soft modes is not observed can be given by the sampling bias due to the gradient descent. The measure with which the minima are sampled is the relative size $\Omega$ of their basins of attraction. Both numerically (employing the Einstein method explained in \cite{xu2011}) and analytically (from a naive reading of the Kac-Rice formula, see for example \cite{auffinger}), we find that $\Omega \sim \det \mat M$, i.e. minima with an abundance of very soft modes would very likely have very small basins of attraction and, thus, not contribute significantly to the empirical distributions.

This observation highlights an important and well-known aspect of the gradient descent: it is an inherently out-of-equilibrium process that should be looked at dynamically. Thus, we are not to consider the DM-ensemble with a permanently non-negative spectrum, but a transition from the initial equilibrium spectrum to a non-negative spectrum under the descent dynamics. Given that the average of the spectrum is set by the Lagrange multiplier $z$, i.e. by the energy, the constantly decreasing energy corresponds to an overall drift in the eigenvalues, shifting them towards higher values. We quantify this by expanding $S_i = S_i + \mathrm{d}\!S_i$ and $z = z +\mathrm{d}z$ in the Hessian \eqref{eq:hessian} to first order, which results in
\begin{align}
\mathrm{d} M_{ij} &= (J_{ijk} + \ldots ) \,\mathrm{d}S + \mathrm{d}z \delta_{ij} \text{.} \label{eq:mat}
\end{align}

Our strategy to make gainful progress from this spectral perspective is to simplify the matrix dynamics of eq.~\eqref{eq:mat} by only considering two important factors that must be there: a source of noise and an entropic confinement establishing a well-defined ensemble. The conceptual background of this approach is the seminal insight by Dyson~\cite{dyson} that, for the Gaussian ensemble, equilibrium sampling in random matrix theory~\cite{mehta2004} can be accomplished by deriving the associated Langevin equation
\begin{align}
\mathrm{d} M^\text{eq}_{ij} &= \mathrm{d}W - \sigma_M^{-2} M^\text{eq}_{ij} \mathrm{d}t. \label{eq:mat2}
\end{align}
Here, the first term is a Gaussian noise term (with $\mathrm{d}W \sim \sqrt{\beta^{-1}\mathrm{d}t}$ being a Wiener process) and the second term is an entropic spring that ensures that the matrix stays in the correct ensemble. This overall structure corresponds to the effective form of that the actual $p$-spin eigenvalue dynamics must have, aside from the energy-dependent drift ($z$-term) of eq.~\eqref{eq:mat}. The rationale for this is to note that the dynamics of the Hessian without the energy-shift are effectively uncorrelated with the change in energy (omitting index permutations) $\mathrm{d}E=J_{ijk} (S_i S_j \mathrm{d} S_{k} + \ldots)$. Thus, we can reduce the first term in eq.~\eqref{eq:mat} to a centered Gaussian increment. The second term in eq.~\ref{eq:mat2} reflects the unavoidable correlations in these updates: there is a well-defined entropic ensemble (the spectrum is a semi-circle at all energies in distribution) and, thus, a restoring drift to this ensemble. Using these increments of the matrix elements, it is straightforward to use matrix perturbation theory to determine the resulting dynamics of the eigenvalues
\begin{align}
\mathrm{d}\lambda_n &= \mathrm{d}W + \mathrm{d}t \left[ -\frac{\lambda_n}{\sigma^2_J} + \sum_{n\neq m} \frac{1}{\lambda_n - \lambda_m} \right].
\end{align}
This is Dyson Brownian motion. Here, $\sigma^2_J$ denotes the $\mathcal{O}(N^{-2})$ variance of the couplings. The repulsive third term with its divergence is the technical equivalent of the statement that two eigenvalues do not cross. Thus, the eigenvalues stay in one representation of the permutation group for all times. 

Comparing the two expressions of eqs~\eqref{eq:mat} and \eqref{eq:mat2}, it is tempting on first glance to regard them as qualitatively identical, since the descent dynamics will also lead to some random noise (which ultimately will be Gaussian due to the Gaussian distribution of  $J$), but with a Hessian that remains in the same well-defined ensemble. The influence of the additional drift term $z \delta_{ij}$ on the eigenvalues might at first seem trivial. However, there is an important difference related to time-translational invariance. In the Dysonian case, everything is in equilibrium and thus invariant in statistics under time-translation and time-reversal. For descent dynamics, these symmetries are trivially broken by the drift. This can easily be handled, but the symmetries are also broken in the noise term. This can be seen if one goes back to earlier analytical approaches~\cite{cugliandolo1993} to the dynamics of the $p$-spin model, where the correlation function $C(t+\Delta t,t)= \langle S_i (t+\Delta t) S_i (t) \rangle$ decays exponentially for $t\gg \Delta t$ with a rate that is inversely proportional to $t$, i.e. we expect
\begin{align}
S_i(t+\Delta t) S_i(t) &= 1- \text{const} \, \Delta t/t + \mathcal{O}\left((\Delta t/t)^2\right) \text{.} 
\end{align}
Thus, even a temporally coarse-grained version of eq.~\eqref{eq:mat} that would get rid of correlations in time without drift would not be Markovian, because the strength of noise would be time-dependent. Since we are only interested in first-passage statistics, this is easily mitigated as we can switch to dynamics in a reparametrized time~\cite{cugliandolo1993} $\tau$ with $\mathrm{d}\tau \propto \mathrm{d}t/t$. This logarithmic reparametrization makes the process time-translational invariant. This freedom to reparametrize time with an arbitrary bijection is very important here. For general systems, we might not a specific procedure to achieve a time-translationally invariant descent, but on a conceptual level this reparametrization vastly constrains the types of essentially different spectral motions, making the Dyson equation of motion as important as it is. One important class of alterations are arising sum rules in the system, as we discuss below in the final section.

The final missing piece is the energy, which decreases with time during gradient descent. We had to separate this drift from the matrix dynamics to bring the latter into a treatable form, so the relevant process is no longer a first crossing of zero, but the first crossing of a curve given by the energy $\varepsilon(\tau)$. As the fluctuations of the energy (given by the gradient, the first derivative of a Gaussian field) are independent from the fluctuations in eigenvalues (aside from the mean they are given by the random part of the Hessian \eqref{eq:hessian}, the second derivative of a Gaussian field) and of lower order ($\mathcal{O}(N^{-2/3}$ in the eigenvalues, but $\mathcal{O}(N^{-1})$ in the intensive energy) we can replace the actual energy by the asymptotic trajectory $\varepsilon_\infty(\tau)$ for large $N$. This of course not only neglects the dynamical fluctuations along the trajectory, but also the deterministic noise in trajectories from the initial conditions. However, we are interested in the behavior at large times $\tau$, where the effect of the initial configurations is negligible.

We can deduce from the structure of the analytical equations~\cite{cugliandolo1993,franz} that the intensive energy in real-time $\varepsilon_\infty(t)\sim \varepsilon_\text{th}+\text{const} t^{-\gamma}$ asymptotically is a power-law, thus in rescaled time the energy decays exponentially $\varepsilon_\infty(\tau)\sim\varepsilon_\text{th}+\text{const} \exp{-\tau/\tau_c}$. A more pedestrian way to look at this is by directly writing down an equation of motion for the energy
\begin{align}
&\partial_t E = \left[\partial_t S_i\right]\left[\partial_{S_i} E\right] \nonumber \\
&= (J_{ijk}J_{ilm} + \ldots) S_j S_k S_l S_m + 3 z J_{ijk} S_i S_j S_k \text{.}
\end{align}
Again changing variables to establish time-translation invariance and performing an average over the disorder we see that $\langle \partial_\tau \varepsilon \rangle = \text{const} + \varepsilon^2$, from which one gathers (using the known asymptotic value) that $\varepsilon_\infty \approx \varepsilon_{th} \operatorname{tanh}(\text{const} \ \tau)$ with the aforementioned asymptotically exponential decay.

We note that this picture is very general and should apply to all of the systems we have studied, not just the $p$-spin model. Putting everything together, we get a model version of the gradient descent as a first-passage process of the lowest eigenvalue of the Hessian, given by standard Dyson dynamics with a time dependent boundary $\varepsilon_\text{model} = \varepsilon_\text{th} \tanh(\tau)$. From this model definition, one can (somewhat {\it a posteriori }admittedly) rationalize the following numerical findings described earlier. (1) It is inherently plausible for the finite-size scaling to be of the same form as the Tracy-Widom distribution because the effective process is indeed one dominated by edge eigenvalue fluctuations; (2) The shape of the distribution (Gaussian on one side and Tracy-Widom on the other) is plausible. The fluctuations of the lowest eigenvalue are of order $N^{-2/3}$ with a hard border to the right, thus there is some time $\tau_0$ where the typical distance is of the same order, so that there is a very high probability for the boundary to not have been crossed. Thus, we know that at $\tau_0$ the fluctuation distribution of the lowest eigenvalue is given by the Tracy-Widom law. However, $\tau_0$ will be close to the actual final time and at small times (up to eigenvalue distances of $\mathcal{O}(N^{-1/2})$) the diffusive part dominates. Thus, it seems within reason that we would see a distribution that effectively looks like a convolution of a Gaussian (the propagator on short times) and the Tracy-Widom law (the fictitious initial condition at $\tau_0$), which would bear the hallmarks we find in the original numerics.

Numerical exploration of this effective description is straightforward with various options for sampling this process. One way would be to go back to the initial idea of the Dyson Brownian motion and diagonalize a matrix subject to small noise in time. The non-crossing is manifest in this approach, but diagonalization is a rather costly operation. Alternatively, one might consider event-based Monte-Carlo of the thermal ensemble whose Langevin equation is given by the Dyson Brownian motion. Finally, there is the option to do straightforward integration of the equation of motion with adaptive time-steps that ensure that the trajectories of eigenvalues never cross. Opting for the latter, we find that we can indeed get to satisfying agreement of the numerics by tuning the strength of the white noise inflicted upon the eigenvalues which one can characterize by an effective temperature $T$. A first-principle determination of the specific value of $T$ to be used is beyond the scope of our arguments. A numerical determination is possible, but simulations of spectral trajectories are slow for two reasons: the cost of the diagonalization itself and the need to switch from FIRE to the direct integration to see the dynamics in physical time. We opt for a more pragmatic procedure and simulate the process for a few values (large values of $T$ are slow as increasing the noise induces more collisions); see Fig.~\ref{fig:fpt} for $N=64$. We see that indeed there is a satisfying agreement between the distributions found from this simple first-passage problem and the real ones for the $p$-spinmodel in Fig.~\ref{fig:combo_pdf}.

\begin{figure}
\includegraphics[width=\linewidth]{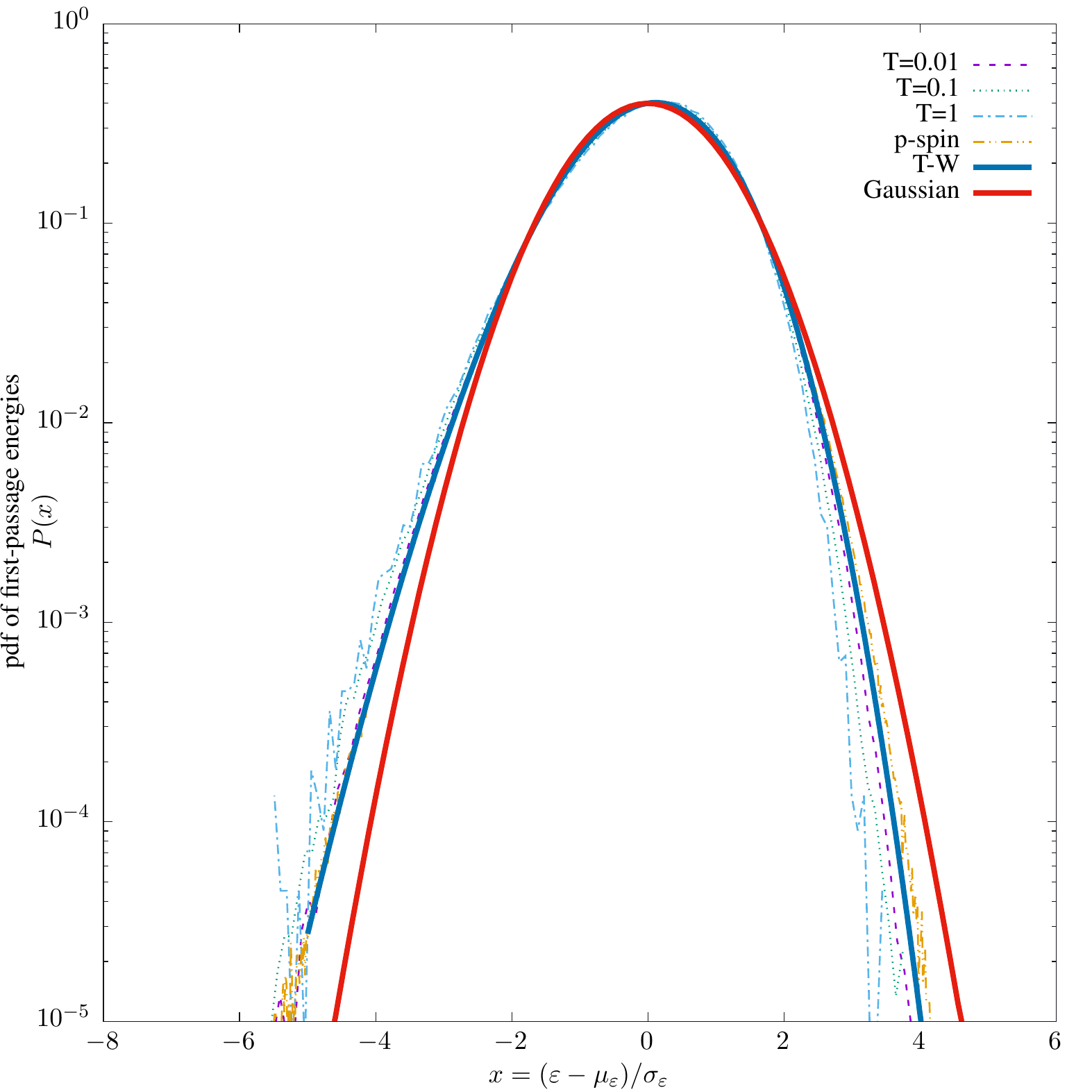}
\caption{Fluctuation distribution as inferred from the first-passage process discussed in the main text for $N=64$ and various values of $T$, together with the actual distribution measured for the $p$-spin model (orange dashed curve). For comparison, we show the Tracy-Widom (blue solid) and Gaussian (red solid) distributions.}
\label{fig:fpt}
\end{figure}

We close with a final remark on the contributions to the final energies that we ignored. Once the lowest eigenvalue has crossed the boundary, one still has descend further to reach the bottom of the minimum. This leads to corrections to the energies of the minima.  All evidence seems to corroborate that these corrections are of higher order in $1/N$ and almost Gaussian distributed. In this case, they do not contribute to the finite-size scaling and also not to the fluctuation distribution of normalized variables as they would only change the first two cumulants. The eigenvalue process outlined here therefore captures the essential mechanism underlying the distribution for the $p$-spin model as well as the distributions for the other constraint satisfaction problems considered here. 

\section{Conclusions}

We performed gradient descent simulations in several prototypical constraint-satisfaction problems with complex landscapes and found similar asymmetric distributions in the normalized distributions of final energies (fig.~\ref{fig:combo_pdf}). These feature a soft tail corresponding to better-than-typical solutions and a hard tail for worse-than-typical solutions. Inspecting in more detail for the spherical $p$-spin spin-glass model, we found that both the finite-size scaling  as well as the functional form of the soft tail (fig.~\ref{fig:fs_distro}) are reminiscent of the Tracy-Widom distribution, which is usually associated with the fluctuations of extremal eigenvalues in random matrix problems.

We made this connection manifest by proposing a novel interpretation of gradient-descent problems as a first-passage process into mechanical stability, i.e. we argue that the energy at which the lowest eigenvalue becomes non-negative is a good proxy for the actual final energy at which the gradient descent terminates with respect to the fluctuation distribution. This is a purely dynamical picture of the out-of-equilibrium gradient-descent process in which typical landscape features such as the basins of attraction are emergent from the random matrix ensemble associated with the dynamical matrix. The very simple nature of the ersatz-process found by reducing the spectral dynamics to their core ingredients could allow for an exact analytical treatment in the future.

An open question remains concerning the extent to which the observed phenomenology survives with increasing system size. At least for the $p$-spin model, extrapolation to the large-$N$ limit does lead to a non-Gaussian distribution with the same tail behavior as seen in finite systems. However, this is less clear for the other models studied here, for which it was difficult to obtain comparable statistics. Nevertheless, the perspective of gradient descent as a first-passage process suggests that the highly similar non-Gaussian features seen in the distributions for the other models are not a finite-size effect, and should persist in the thermodynamic limit.

The view of the gradient descent process as a first-passage problem could be a rather broadly fruitful one. Most aspects of the (matrix) dynamics of the $p$-spin model are believed to be somewhat general for many complex systems. Additionally, the topological feature that fluctuations towards lower energies (corresponding to minima with atypically soft modes) are substantially easier to find than those towards higher energies (hard modes) should prevail in a vast variety of systems with complex landscapes. This way of thinking should be helpful in understanding phenomenology in experiments such as Ref.~\citenum{nakagawa} that prominently feature asymmetrical distributions of the fluctuations within the inherent structure landscape. Our reasoning should be applicable to results from finite temperature quenches as long as the initial temperature is sufficiently high that the system is ergodic and the final temperature sufficiently low that the system is confined to a single basin after the quench.  Finally, we note that a good understanding of the first passage into mechanical stability might inspire new ways of tweaking interactions to convert complex landscapes into less rough ones  (similar to the methods proposed in ref.~\citenum{ruiz2019}) in order to find better (lower energy) solutions. 

We end with a caveat: In finite-dimensional models and data, a simple Dysonian random-matrix view as proposed here will necessarily face some issues, one very important one being the existence of sum rules constraining the Hessian, particularly the ones corresponding to mechanical equilibrium. The details of the coordination structure have been argued\cite{manning2015,stanifer2018,benetti2018} to be crucial in understanding essential features of low-dimensional jammed packings such as the scaling of the vibrational density of states. As this is directly linked to the statistics of the extremal eigenvalues, it is very intriguing for future work to study the effect of these constraints (which develop as the system descends in the landscape) on the distributions studied here. Even when such constraints exist, however, the notion is still valid that there is one contributing process to the statistics of descents in disordered landscapes, related to the passage into mechanical stability that we have isolated for the $p$-spin model.

\acknowledgments 
We thank S. Teitel for providing us with the data used in Ref.~\citenum{vagberg}, S. Ridout for additional isobaric jammed configurations and S. R. Nagel for discussions that inspired this investigation.
This work was supported by the "Cracking the glass problem" collaboration grant  (348126 to HHB, 454943 to JK and 454945 to AJL), as well as an Investigator grant (327939 to AJL) from the Simons Foundation.
\appendix

\end{document}